\documentclass[%
reprint,
superscriptaddress,
 amsmath,amssymb,amsfonts,
 aps,
pra,
]{revtex4-1}

\usepackage[english]{babel}
\usepackage[utf8x]{inputenc}
\usepackage[T1]{fontenc}

\usepackage{amsmath}
\usepackage{graphicx}
\usepackage[colorinlistoftodos]{todonotes}
\usepackage[colorlinks=true, allcolors=blue]{hyperref}
\usepackage{graphicx,amssymb,amsmath,amsthm,mathtools,mathrsfs,hyperref}
\usepackage{graphicx}

\theoremstyle{definition}

\usepackage{dcolumn}
\usepackage{bm}

\begin{document}

\preprint{APS/123-QED}

\title{Algebraic approach for investigation of a multi-mode quantum system dynamics}

\author{Andrei Gaidash}%
\email{andrei_gaidash@itmo.ru}
\affiliation{Department of Mathematical Methods for Quantum
  Technologies, Steklov Mathematical Institute of Russian Academy of
  Sciences, 119991, 8 Gubkina St, Moscow, Russia} 
\affiliation{Laboratory of Quantum Processes and Measurements, ITMO University, 199034, 3b Kadetskaya Line, Saint Petersburg, Russia}

\author{Anton Kozubov}
 \affiliation{Department of Mathematical Methods for Quantum
  Technologies, Steklov Mathematical Institute of Russian Academy of
  Sciences, 119991, 8 Gubkina St, Moscow, Russia}
\affiliation{Laboratory of Quantum Processes and Measurements, ITMO University, 199034, 3b Kadetskaya Line, Saint Petersburg, Russia}
  
\author{Alexei Kiselev}
\affiliation{Laboratory of Quantum Processes and Measurements, ITMO University, 199034, 3b Kadetskaya Line, Saint Petersburg, Russia}
\affiliation{Faculty of Photonics and Optoinformatics, ITMO
  University, 199034, 3b Kadetskaya Line, Saint Petersburg, Russia}
\affiliation{Faculty of Physics, St.\,Petersburg State University, 199034, Saint Petersburg, Russia}

\author{George Miroshnichenko}
\affiliation{Waveguide Photonics Research Center, ITMO University, 197101, 49 Kronverksky Pr., Saint Petersburg, Russia}
\affiliation{Institute <<High School of Engineering>>, ITMO University, 197101, 49 Kronverksky Pr., Saint Petersburg, Russia}%

\date{\today}

\begin{abstract}
We introduce algebraic approach for superoperators that might be useful tool for investigation of quantum (bosonic) multi-mode systems and its dynamics. In order to demonstrate potential of proposed method we consider multi-mode Liouvillian superoperator that describes relaxation dynamics of a quantum system (including thermalization and intermode coupling). Considered algebraic structure of superoperators that form Liouvillian and their algebraic properties allows us to diagonilize multi-mode Liouvillian to find its spectrum. Also it allows to derive linear by mean number of thermal (environmental) photons approximation for time-evolution superoperator that keeps amount of considered dimensions in Fock space finite (assuming initial amount of dimensions finite) that might be helpful regarding entanglement dynamics problems. Conjugate Liouvillian is considered as well in order to perform analysis in Heisenberg picture, it can be implemented for multi-time correlation functions derivation.
\end{abstract}

\maketitle


\section{Introduction}
There are lots of physical approaches to study quantum channels, properties of quantum states during propagation and concomitant optical effects. e.g. \cite{lindblad1975completely, carmichael2009open, omnes1997general, lidar2003decoherence, zurek2003decoherence, miroshnichenko2012hamiltonian, miroshnichenko2018decoherence, rivas20132, kozubov2019quantum,gaidash2020dissipative,gaidash2021quantum}. Authors utilize combination of the effective Hamiltonian derived in~\cite{miroshnichenko2012hamiltonian} and a Lindbland-like master equation~\cite{lindblad1976generators, gorini1976completely} that lies at the heart of the theoretical approach to the quantum dynamics of optical quantum states in~\cite{miroshnichenko2018decoherence,kozubov2019quantum,gaidash2020dissipative,gaidash2021quantum}.

In \cite{gaidash2020dissipative} algebraic approach based on $SU(1,1)$ was proposed in order to diagonalize the Liouvillian and find evolution of density matrix in linear by mean number of thermal photons approximation, however environment-mediated intermode coupling was not considered. In \cite{gaidash2021quantum} the latter effect was considered but only regarding dynamics of mean values of quadratic operators, evolution of density matrix was out of scope. However evolution of density matrix rather than mean values of some operators may be useful considering some problems, for instance dynamics of observables of general form or dynamics of entanglement of Fock states (e.g. Bell states) \cite{bertlmann2002geometric,bertlmann2005optimal}. The article continues series of papers \cite{miroshnichenko2018decoherence,kozubov2019quantum,gaidash2020dissipative,gaidash2021quantum} considering the latter problem.

The challenge regarding finding of direct solution for evolution of density matrix in case of multi-mode Liovillian with intermode coupling was caused by presence of mixed algebras. For instance if we consider two-mode system then each mode can be described in terms of $SU(1,1)$ algebra (losses) while effects of intermode coupling can be described by $SU(2)$ algebra (rotations). Mixed algebras are hard to deal with and often require some additional efforts, e.g. introduction of analysis similar to supersymmetric one \cite{fayet1977supersymmetry,sohnius1985introducing}.

In order to deal with it in this paper we introduce superoperator algebra that resembles Jordan-Schwinger map for quadratic operators \cite{jordan1935zusammenhang,schwinger2001angular}, one may observe how matrix of numerical coefficients alters due to some operations (e.g. commutation) with superoperators. In particular developed algebra allows to diagonalize Liouvillian. Diagonalization is necessary for finding spectrum of the multi-mode Liouvillian and also it allows to find linear by mean number of thermal photons approximation. The latter is essential for analysis, including numerical estimations, since it allows to operate with limited amount of dimensions in Fock space (assuming initial state has finite dimensions), i.e. considering $m$-mode system where each mode may contain up to $n$ photons then the system has $(n+1)^m$ dimensions while linear approximation increases number of dimensions only to $(n+2)^m$ compare to infinite number of dimensions for full solution.

The article is organized as follows. In Section~\ref{problem} we consider multi-mode Liouvillian and introduce the superoperators. In Section~\ref{algebra} we introduce algebra and investigate its properties. Section~\ref{diagonal} is dedicated to diagonalization of Liouvillian as well as to its corollaries. Section~\ref{conclusion} concludes the paper.

\section{Preliminaries}\label{problem}
We would like to introduce following multi-mode Liouvillian (master equation in generalized Gorini-Kossakowski-Sudarshan-Lindblad
(GKSL) form \cite{lindblad1976generators, gorini1976completely}) expressed in superoperator form \cite{gaidash2020dissipative}:
\begin{gather}
    \hat{L}=\nonumber\\
    =\sum_{n,m}\Bigg[-i\Omega_{nm}\hat{n}^{(-)}_{nm}-\Gamma_{nm}\Big((2n_T+1)\hat{K}_{nm}^{(0)}-\nonumber\\
    -n_T\hat{K}_{nm}^{(+)}
    -(n_T+1)\hat{K}_{nm}^{(-)}-\frac12\delta_{nm}\Big)\Bigg],\label{L1}
\end{gather}

where $\Omega$ and $\Gamma$ are frequency and relaxation matrices correspondingly and
\begin{gather}
    \hat{n}^{(-)}_{nm}=\overleftarrow{\hat{a}^\dagger_n \hat{a}_m}-\overrightarrow{\hat{a}^\dagger_n \hat{a}_m},
\end{gather}
is Hamiltonian related term that is described by the Liouville notation (i.e. $\overleftarrow{A}B=AB$ and $\overrightarrow{A}B=BA$ \cite{vadeuiko2000diagonal}), terms related to relaxation superoperator are described by the Liouville notation as well:
\begin{gather}
    \hat{K}_{nm}^{(+)}=\overleftarrow{\hat{a}^\dagger_n} \overrightarrow{\hat{a}_m},\\
    \hat{K}_{nm}^{(-)}=\overleftarrow{\hat{a}_m}\overrightarrow{\hat{a}^\dagger_n} ,\\
    \hat{K}_{nm}^{(0)}=\frac{1}{2}\left(\overleftarrow{\hat{a}^\dagger_n\hat{a}_m} +\overrightarrow{\hat{a}_m\hat{a}^\dagger_n}\right),
\end{gather}
where $\hat{a}_j$ ($\hat{a}^\dagger_j$) is annihilation (creation) operator of $j^{th}$ mode, $n_T$ is mean number of environmental thermal photons, $\delta_{ij}$ is Kronecker symbol. These four types of superoperators $\hat{n}^{(-)}_{nm}$, $\hat{K}_{nm}^{(+)}$, $\hat{K}_{nm}^{(-)}$, and $\hat{K}_{nm}^{(0)}$ generate an algebra. In the following section we introduce notation that provides reduced number of commutation relations as well as more clear picture of its structure.

\section{Algebraic approach}\label{algebra}
In this section we introduce algebra of superoperators that resembles Jordan-Schwinger approach for operators: commutation relations are applied not only to superoperators itself but also to their matrix coefficients. Basic algebraic properties are described further and their application to the problem of Liouville diagonalization is considered in the next Section.

\subsection{Superoperator with matrix coefficient notation}
Let us introduce following notation:

\begin{align}
    \hat{\mathcal{P}}_{A}
    \equiv \sum_{n,m} A_{nm} \hat{P}_{nm},
\end{align}
where $\hat{P}_{nm}$ is an arbitrary (super)operator that has two indices (presumably quadratic), $A_{nm}$ is complex coefficient. Then according to the introduced notation (super)operators have the following properties:
\begin{gather}
    \hat{\mathcal{P}}_{A}+\hat{\mathcal{Q}}_{A}=(\hat{\mathcal{P}}+\hat{\mathcal{Q}})_{A},\\
    \hat{\mathcal{P}}_{A}+\hat{\mathcal{P}}_{B}=\hat{\mathcal{P}}_{A+B},\\
    c\cdot\hat{\mathcal{P}}_{A}=\hat{\mathcal{P}}_{c\cdot A}.
\end{gather}
Also one may observe following useful property, if 
\begin{gather}
    [\hat{P}_{ij},\hat{Q}_{nm}]=\hat{S}_{im}\delta_{jn}\pm\hat{S}_{jn}\delta_{im},\label{prop1}
\end{gather}
where $[A,B]=AB-BA$ is commutator, then
\begin{gather}
    [\hat{\mathcal{P}}_{A},\hat{\mathcal{Q}}_{B}]=\hat{\mathcal{S}}_{AB\pm BA}.\label{prop2}
\end{gather}
The latter expression fully recovers basic principles of Jordan-Schwinger map if $\hat{P}_{ij}$, $\hat{Q}_{ij}$ and $\hat{S}_{ij}$ are quadratic operators. However in case of superoperators algebra becomes more peculiar; one have an opportunity to see this in the following subsections.

One may express alternative form of the Liouvillian as follows:
\begin{gather}
    \hat{L}=\hat{\mathcal{N}}^{(-)}_{-i\Omega}-\hat{\mathcal{K}}^{(0)}_{(2n_T+1)\Gamma}
    + \hat{\mathcal{K}}^{(+)}_{n_T\Gamma}+\hat{\mathcal{K}}^{(-)}_{(n_T+1)\Gamma}+\mathcal{I}_{\frac12\Gamma},
    \label{altL}\\
    \mathcal{I}_{A}=\sum_{n,m}A_{nm}\delta_{nm}=\text{Tr}(A),
\end{gather}
where $\text{Tr}(\ \cdot\ )$ is trace operation. In the following subsection we discuss basic commutation properties of considered superoperators that form Liouvillian.

\subsection{Commutation relations}
One of the most crucial investigation tools for an algebra is to consider commutation relation of an operators. Let us have a closer look on $\hat{n}^{(-)}_{nm}$, $\hat{K}_{nm}^{(+)}$, $\hat{K}_{nm}^{(-)}$, and $\hat{K}_{nm}^{(0)}$ and on how one may use introduced earlier notation with matrix coefficients. Commutation relations are as follows:
\begin{gather}
    [\hat{K}^{(0)}_{ij},\hat{K}^{(\pm)}_{nm}]=\pm\frac12(\hat{K}^{(\pm)}_{im}\delta_{jn}+\hat{K}^{(\pm)}_{jn}\delta_{im}),\\
    [\hat{K}^{(-)}_{ij},\hat{K}^{(+)}_{nm}]=\hat{K}^{(0)}_{im}\delta_{jn}+\hat{K}^{(0)}_{jn}\delta_{im}-\nonumber\\
    -\frac12\big(\hat{n}^{(-)}_{im}\delta_{jn}-\hat{n}^{(-)}_{jn}\delta_{im}\big),\\
    [\hat{n}^{(-)}_{ij}, \hat{K}^{(s)}_{nm}]=\hat{K}^{(s)}_{im}\delta_{jn}-\hat{K}^{(s)}_{jn}\delta_{im},
\end{gather}
where $s$ is either <<$0$>>, <<$+$>> or <<$-$>>. Then the following is straightforward by utilizing the property described in Eqs.~\ref{prop1} and~\ref{prop2}:
\begin{gather}
    [\hat{\mathcal{K}}^{(0)}_{A}, \hat{\mathcal{K}}^{(\pm)}_{B}]
     =\pm\hat{\mathcal{K}}^{(\pm)}_{\frac12\{A,B\}},\label{algebra1}\\
    [\hat{\mathcal{K}}^{(-)}_{A}, \hat{\mathcal{K}}^{(+)}_{B}]
     =\hat{\mathcal{K}}^{(0)}_{\{A,B\}}-\hat{\mathcal{N}}^{(-)}_{\frac12[A,B]},\label{algebra2}\\
     [\hat{\mathcal{N}}^{(-)}_{A}, \hat{\mathcal{K}}^{(s)}_{B}]=\hat{\mathcal{K}}^{(s)}_{[A,B]},\label{algebra3}\\
     [\hat{\mathcal{K}}^{(\pm)}_{A}, \hat{\mathcal{K}}^{(\pm)}_{B}]=0,\label{algebra4}
\end{gather}
where $\{A,B\}=AB+BA$ is anticommutator.

Algebraic structure is described in Eqs.~\ref{algebra1} -- \ref{algebra4} resembles $SU(1,1)$ and may be reduced to it by a restrictions on matrix coefficients (the easiest way is to consider not matrices but numbers as coefficients, i.e. $1 \times 1$ matrix). However in our case matrices $\Omega$ and $\Gamma$ are arbitrary (both are Hermitian and $\Gamma>0$) so further we should consider general case.

\subsection{Orthogonal transformation}
As in \cite{gaidash2020dissipative} one should apply exponentiated superoperators in order to diaginalize Liouvillian. However at first one should consider following application type of Baker-Campbell-Huasdorff expression (these are corollary from commutation relations from the previous subsection):

\begin{align}
    e^{\hat{\mathcal{K}}^{(\pm)}_{B}} \hat{\mathcal{N}}^{(-)}_{A} e^{-\hat{\mathcal{K}}^{(\pm)}_{B}} &= \hat{\mathcal{N}}^{(-)}_{A} - \hat{\mathcal{K}}^{(\pm)}_{[A,B]},\label{orth1}\\
    e^{\hat{\mathcal{K}}^{(\pm)}_{B}} \hat{\mathcal{K}}^{(0)}_{A} e^{-\hat{\mathcal{K}}^{(\pm)}_{B}} &= \hat{\mathcal{K}}^{(0)}_{A}\mp\hat{\mathcal{K}}^{(\pm)}_{\frac12\{A,B\}},\\
    e^{\hat{\mathcal{K}}^{(\pm)}_{B}} \hat{\mathcal{K}}^{(\mp)}_{A} e^{-\hat{\mathcal{K}}^{(\pm)}_{B}} &= \hat{\mathcal{K}}^{(\mp)}_{A}\mp\hat{\mathcal{K}}^{(0)}_{\{A,B\}}+\hat{\mathcal{N}}^{(-)}_{\frac12[A,B]}+\hat{\mathcal{K}}^{(\pm)}_{BAB}\label{orth2}.
\end{align}
The latter expressions also demonstrate structure of introduced algebra from a different point of view; the latter property clearly shows that algebra is closed.

Then one should apply the latter orthogonal transformations to Liouvillian as follows:
\begin{gather}
    e^{\hat{\mathcal{K}}^{(+)}_{B}} \hat{L} e^{-\hat{\mathcal{K}}^{(+)}_{B}}=\nonumber\\
    = \hat{\mathcal{N}}^{(-)}_{P}-\hat{\mathcal{K}}^{(0)}_{Q}
    +\hat{\mathcal{K}}^{(+)}_{R}+\hat{\mathcal{K}}^{(-)}_{(n_T+1)\Gamma}+\mathcal{I}_{\frac12\Gamma},\\
    P=-i\Omega+\frac12(n_T+1)[\Gamma,B],\\
    Q=(2n_T+1)\Gamma+(n_T+1)\{\Gamma,B\},\\
    R=i[\Omega,B]+\frac12(2n_T+1)\{\Gamma,B\}+n_T\Gamma+(n_T+1)B\Gamma B,\label{ricatti1}
\end{gather}
and
\begin{gather}
    e^{\hat{\mathcal{K}}^{(-)}_{A}} e^{\hat{\mathcal{K}}^{(+)}_{B}} \hat{L} e^{-\hat{\mathcal{K}}^{(+)}_{B}} e^{-\hat{\mathcal{K}}^{(-)}_{A}}=\nonumber\\
=\hat{\mathcal{N}}^{(-)}_{X}+\hat{\mathcal{K}}^{(0)}_{Y}+\hat{\mathcal{K}}^{(+)}_{R}+\hat{\mathcal{K}}^{(-)}_{Z}+\mathcal{I}_{\frac12\Gamma},\label{diageq}\\
X=P+\frac12[R,A],\\
Y=\{R,A\}-Q,\\
Z=[A,P]-\frac12\{Q,A\}+ARA+(n_T+1)\Gamma.
\end{gather}

\section{Liouvillian diagonalization}\label{diagonal}

In order to diagonalize Liouvillian one should find matrices $A$ and $B$ that nullify matrix indices for $\hat{\mathcal{K}}^{(+)}$ and $\hat{\mathcal{K}}^{(-)}$ in Eq.~\ref{diageq}, i.e. $R=Z=0$.
The latter equations are algebraic Ricatti equations that have no explicit solution in general case. However we have found solution with the help of observations shown in Apps.~\ref{app1} and~\ref{app2}, it is as follows:
\begin{gather}
   R=0 \quad \rightarrow \quad B=-\frac{n_T}{n_T+1}I,\\
    Z=0 \quad \rightarrow \quad A=(n_T+1)I,
\end{gather}
where $I$ is identity matrix.
Then
    \begin{gather}
    e^{\hat{\mathcal{K}}^{(-)}_{A}} e^{\hat{\mathcal{K}}^{(+)}_{B}} \hat{L} e^{-\hat{\mathcal{K}}^{(+)}_{B}} e^{-\hat{\mathcal{K}}^{(-)}_{A}}=\hat{L}^{(d)},\label{ld}\\
    \hat{L}^{(d)}=\hat{\mathcal{N}}^{(-)}_{-i\Omega}+\hat{\mathcal{K}}^{(0)}_{-\Gamma}+\mathcal{I}_{\frac12\Gamma},
\end{gather}
where $\hat{L}^{(d)}$ is diagonalized Liouvillian, and
    \begin{gather}
    \hat{L}=e^{-\hat{\mathcal{K}}^{(+)}_{B}} e^{-\hat{\mathcal{K}}^{(-)}_{A}} \hat{L}^{(d)} e^{\hat{\mathcal{K}}^{(-)}_{A}} e^{\hat{\mathcal{K}}^{(+)}_{B}},\label{l1}
\end{gather}
also keeping in mind that $\hat{\rho}(t)=e^{\hat{L}t}\hat{\rho}_0$.

\subsection{Spectrum of diagonalized Liouvillian}\label{subsec-spectrum}
Spectrum of diagonalized Liouvillian can be found in the following form:
\begin{gather}
    \hat{L}^{(d)}\hat{\rho}=\lambda \hat{\rho}.
\end{gather}
Finding spectrum of a (super)operator is essential for investigation of its properties, especially when it describes dynamics of a quantum state. For instance, spectrum of (diagonalized) Liouvillian is closely related to ``velocities'' or ``frequencies'' of the main processes that contribute to the time evolution; introduced orthogonal transformations do not change these values.

Further we consider the case of two-mode Liouvillian for the sake of visual simplicity (however it is also one of the most important case that describes dynamics related to polarization modes), it may be generalized by following the same steps as it is shown in more details in App.~\ref{appeigen}. The result is as follows:
  \begin{gather}
      \sum_{a,b} C_{a,b}^{u,v} (\hat{L}^{(d)})_{(a, n),(b,m)}^{u,v}= \lambda^{u,v}C_{n,m}^{u,v},
  \end{gather}
where
\begin{gather}
    (\hat{L}^{(d)})_{(a, n),(b, m)}^{u,v}=\nonumber\\
    =\big(A_{11}n+B_{11}m+A_{22}(u-n)
    +B_{22}(v-m)\big)\delta_{a,n}\delta_{b,m}+\nonumber\\
    A_{12}\sqrt{n(u-n+1)}\delta_{a,n-1}\delta_{b,m}+\nonumber\\
    +B_{12}\sqrt{(m+1)(v-m)}\delta_{a,n}\delta_{b,m+1}+\nonumber\\
    +A_{21}\sqrt{(n+1)(u-n)}\delta_{a,n+1}\delta_{b,m}+\nonumber\\
    +B_{21}\sqrt{m(v-m+1)}\delta_{a,n}\delta_{b,m-1},\label{lmatrix}
\end{gather}
where
\begin{gather}
    A_{nm}=-i\Omega_{nm}-\frac{\Gamma_{nm}}{2},\\
    B_{nm}=i\Omega_{nm}-\frac{\Gamma_{nm}}{2}.
\end{gather}

Obtained expression for $(\hat{L}^{(d)})_{(a, n),(b, m)}^{u,v}$ is tensor. However one may reduce it to the form of a matrix (by introducing new indices, e.g. by introducing tensor product, i.e. $|a\rangle\langle b| \rightarrow |a\rangle \otimes|b\rangle$) for simpler analysis. As a matrix it has five-diagonal form that has no explicit solution in general, however one may perform numerical estimations easily for a given (polarization and/or spatial) mode structure. Increase of number of interacting modes to the total number of $j$ as 
\begin{gather}
 (\hat{L}^{(d)})_{(a,n),(b,m)}^{u,v}\rightarrow(\hat{L}^{(d)})_{(a,n),(b,m),(c,p),...}^{u,v,w,...}   
\end{gather}
corresponds to consideration of $(2j+1)$-diagonal matrix utilizing the same procedure.

\subsection{Linear Approximation}
As it was discussed earlier linear by the mean number of thermal photons approximation is useful for analysis since it keeps the number of dimensions in Fock space finite. Diagonalized Liouvillian can be complemented by the zero order one (only relaxation, preserves number of initial dimensions) that can be introduced as follows:
\begin{gather}
    \hat{L}^{(0)}=e^{-\hat{\mathcal{K}}^{(-)}_{I}} \hat{L}^{(d)} e^{\hat{\mathcal{K}}^{(-)}_{I}}=\nonumber\\
    =\hat{\mathcal{N}}^{(-)}_{-i\Omega}+\hat{\mathcal{K}}^{(0)}_{-\Gamma}+\hat{\mathcal{K}}^{(-)}_{\Gamma}+\mathcal{I}_{\frac12\Gamma}\label{lo},
\end{gather}
and
\begin{gather}
    e^{\hat{L}t}=e^{\frac{n_T}{n_T+1}\hat{\mathcal{K}}^{(+)}_{I}} e^{-n_T\hat{\mathcal{K}}^{(-)}_{I}} e^{\hat{L}^{(0)}t} e^{n_T\hat{\mathcal{K}}^{(-)}_{I}} e^{-\frac{n_T}{n_T+1}\hat{\mathcal{K}}^{(+)}_{I}}\label{llo}.
\end{gather}

Then we would like to find linear in $n_T$ solution with the help of Baker-Campbell-Hausdorff expression (applied twice):
\begin{gather}
    e^{a\hat{X}}e^{b\hat{Z}}\hat{Y}e^{b\hat{Z}}e^{-a\hat{X}}=\hat{Y}+a[\hat{X},\hat{Y}]+b[\hat{Z},\hat{Y}]+...,\\
    e^{\hat{L}t}\approx e^{\hat{L}^{(0)}t}+n_T[\hat{\mathcal{K}}^{(+)}_{I},e^{\hat{L}^{(0)}t}]-n_T[\hat{\mathcal{K}}^{(-)}_{I},e^{\hat{L}^{(0)}t}].
\end{gather}
Detailed derivation of the following result is in App.~\ref{appder}:
\begin{gather}
    e^{\hat{L}t}\approx\Big(I+n_T(\hat{\mathcal{K}}^{(+)}+2\hat{\mathcal{K}}^{(0)}+\hat{\mathcal{K}}^{(-)})_{I-U(t)}\Big)e^{\hat{L}^{(0)}t},\label{linearres}
\end{gather}
where 
\begin{gather}
    U(t)=e^{(-i\Omega-\frac{\Gamma}{2})t}\cdot e^{(i\Omega-\frac{\Gamma}{2})t}.
\end{gather}
Obtained approximation in Eq.~\ref{linearres} is agreed with the one from \cite{gaidash2020dissipative}, it is generalized on the case of multiple interacting modes.

\subsection{Conjugate Liouvillian}

It may be useful to consider alternative view on the problem regarding dynamics. Considered above approach as well as in previous works \cite{miroshnichenko2018decoherence,kozubov2019quantum,gaidash2020dissipative,gaidash2021quantum} are in Schrodinger picture (density matrix dependent on time). Here we also would like to consider conjugated Liouvillian in order to describe dynamics in Heisenberg picture (operators are dependent on time) as follows by utilization of trace properties:
\begin{gather}
    \langle \hat{A}\rangle(t)=\text{Tr}(\hat{A}\hat{\rho}(t))=\text{Tr}(\hat{A}e^{\hat{L}t}\hat{\rho}_0)\rightarrow\nonumber\\ \rightarrow\text{Tr}(e^{\hat{L}^{+}t}\hat{A}\hat{\rho}_0)=\text{Tr}(\hat{A}(t)\hat{\rho}_0)=\langle \hat{A}(t)\rangle.
\end{gather}

So for conjugate Liovillian basically all ``left'' ($\overleftarrow{A}$) superoperators in Liouville notation become ``right'' ($\overrightarrow{A}$) and vise versa. The latter leads to interchange of $\hat{\mathcal{K}}^{(+)}$ and $\hat{\mathcal{K}}^{(-)}$ as well as complex conjugate of matrix indices:
    \begin{gather}
    \hat{L}^+=\hat{\mathcal{N}}^{(-)}_{i\Omega}-\hat{\mathcal{K}}^{(0)}_{(2n_T+1)\Gamma}
    +\hat{\mathcal{K}}^{(+)}_{(n_T+1)\Gamma}+\hat{\mathcal{K}}^{(-)}_{n_T\Gamma}+\mathcal{I}_{\frac12\Gamma}\label{conjl}.
\end{gather}
All described above (diagonalization and linear approximation) can be applied to the conjugate Liouvillian in the same way keeping in mind the interchange.

\section{Results and discussion}\label{conclusion}

In this paper we introduce algebraic approach to superoperators that resembles Jordan-Schwinger approach: set of superoperators is closed and algebraic operations on them affect their matrix indices. In general, space of superoperators is isomorphic to space of tensor products, potentially one may reduce some superoperator related problem to algebraic polynomial problem and then investigate symmetry of a multi-mode quantum system and its dynamics. Then it is straightforward that introduced algebraic approach may have wide implementation regarding mentioned topics.

In particular, introduced algebraic approach allows to deal with peculiarities of multi-mode Liouvillian diagonalization in order to investigate its properties in details; this paper is dedicated to this specific problem. Multi-mode Liovillian can be expressed in terms of four superoperators $\hat{\mathcal{N}}^{(-)}$, $\hat{\mathcal{K}}^{(+)}$, $\hat{\mathcal{K}}^{(-)}$, and $\hat{\mathcal{K}}^{(0)}$ (with corresponding matrix subscripts) as in Eq.~\ref{altL} and their algebraic structure is described in Eqs.~\ref{algebra1} -- \ref{algebra4}.

Introduced algebraic approach allows us to investigate spectrum of considered multi-mode Liouvillian. We consider orthogonal transformations as in Eqs.~\ref{orth1} -- \ref{orth2} that results to Eq.~\ref{diageq}. The latter is reduced to two algebraic Riccati equations that have been solved in order to diagonalize Liouvillian as in Eq.~\ref{ld}. It is also should be mentioned that relaxation-only Liovillian (zero order by mean number of thermal photons or considering zero temperature of the environment) can be connected with full Liouvilian as well as to its diagonal counterpart by the same orthogonal transformations as in Eqs.~\ref{lo} and~\ref{llo}. Thus all of them have the same spectrum and it can be found by following the steps that are described in Sec.~\ref{subsec-spectrum} and App.~\ref{appeigen}; its eigenvalues corresponds to $(2j+1)$-diagonal matrix, where $j$ is total number of interacting modes. Essential case of two (polarization) modes is considered, however it can be easily generalized to arbitrary number of interacting modes, e.g. in order to investigate dynamics of light in twisted fibers as in \cite{vavulin2017robust}.

Another corollary is linear by mean number of thermal photons approximation of the evolution superoperator in Eq.~\ref{linearres}. Dynamics of entanglement for Gaussian type of states (sqeezed states, Schrodinger cat and etc.) can be considered in framework presented in \cite{kiselev2021lindblad}. However general approach for criterion of entanglement (including Fock states) requires the nearest separable state (e.g. see lemma in \cite{bertlmann2005optimal}), from our point of view it is considerably easier to search for the nearest separable state when dimensions in Fock space are finite (assuming initial state has finite dimensions as well), i.e. considering $m$-mode system where each mode may contain up to $n$ photons then the system has $(n+1)^m$ dimensions while linear approximation increases number of dimensions only to $(n+2)^m$.

The last corollary is derivation of conjugated Liouvillian as in Eq.~\ref{conjl}. This allows to consider dynamics of operators in Heisenberg picture. In particular it might be useful regarding investigation multiple-time correlation functions dynamics and how parameters of a channel may affect on them.

Described in this paper particular implementations are mostly to demonstrate the potential of introduced algebraic approach. At the end of the day we believe that it can be useful tool for future researches in various fields related to physics of quantum systems dynamics. 


\section*{Acknowledgements}
The work of A. Gaidash, A. Kiselev and A. Kozubov is financially supported by the Russian Ministry of Education (Grant No. 2019-0903). 

\appendix
\section{Diagonalization}\label{app1}
Let us consider diagonalization of Liouvillian in single-mode case in details. Then
\begin{gather}
    e^{\beta\hat{K}^{(+)}}\hat{K}^{(0)}e^{-\beta\hat{K}^{(+)}}=\hat{K}^{(0)}-\beta \hat{K}^{(+)},\\
    e^{\beta\hat{K}^{(+)}}\hat{K}^{(-)}e^{-\beta\hat{K}^{(+)}}=\hat{K}^{(-)}-2\beta\hat{K}^{(0)}+\beta^2 \hat{K}^{(+)},\\
    e^{\alpha\hat{K}^{(-)}}\hat{K}^{(0)}e^{-\alpha\hat{K}^{(-)}}=\hat{K}^{(0)}+\alpha \hat{K}^{(-)},\\
    e^{\alpha\hat{K}^{(-)}}\hat{K}^{(+)}e^{-\alpha\hat{K}^{(-)}}=\hat{K}^{(+)}+2\alpha\hat{K}^{(0)}+\alpha^2 \hat{K}^{(-)},
\end{gather}
and
\begin{gather}
    e^{\alpha\hat{K}^{(-)}}e^{\beta\hat{K}^{(+)}}(a\hat{K}^{(0)}+b\hat{K}^{(+)}+c\hat{K}^{(-)})e^{-\beta\hat{K}^{(+)}}e^{-\alpha\hat{K}^{(-)}}=\nonumber\\
    =\hat{K}^{(0)}\Big(a+2(b\alpha-c\beta-a\alpha\beta+c\alpha\beta^2)\Big)+\nonumber\\
    +\hat{K}^{(+)}\Big(b+c\beta^2-a\beta\Big)+\nonumber\\
    +\hat{K}^{(-)}\Big(c+a\alpha+b\alpha^2-a\alpha^2\beta-2c\alpha\beta+c\alpha^2\beta^2\Big).
\end{gather}
In order to nullify coefficients of $\hat{K}^{(+)}$ and $\hat{K}^{(-)}$ one may find following solutions for $\alpha$ and $\beta$:
\begin{gather}
    c\beta^2-a\beta+b=0,\\
    (a-2c\beta)\alpha+c=0.
\end{gather}
In our case $a=(2n_T+1)$, $b=-n_T$ and $c=-(n_T+1)$ hence we have two solutions as follows:
\begin{gather}
     \alpha=-(n_T+1) \quad\text{and}\quad \beta=-1,
\end{gather}
which simplifies 
\begin{gather}
    a\hat{K}^{(0)}+b\hat{K}^{(+)}
    +c\hat{K}^{(-)}\rightarrow -\hat{K}^{(0)},
\end{gather}
and
\begin{gather}
     \alpha=n_T+1 \quad\text{and}\quad \beta=-\frac{n_T}{n_T+1},
\end{gather}
which simplifies 
\begin{gather}
    a\hat{K}^{(0)}+b\hat{K}^{(+)}
    +c\hat{K}^{(-)}\rightarrow \hat{K}^{(0)}.
\end{gather}
The latter case is valid from the physical point of view (negative sign in the former case leads to increase of energy, i.e. pump case, and we consider relaxation).

\section{Ricatti equation}\label{app2}
Let us consider linear by $n_T$ solutions of Eq.~\ref{ricatti1} as $B
\approx B_0+n_T B_1$ that may simplify the equation. Then the zero order equation is as follows:
\begin{gather}
   (i\Omega+\frac12\Gamma)B_0+B_0(-i\Omega+\frac12\Gamma)+B_0\Gamma B_0=0,
\end{gather}
that has obvious solution $B_0=0$, and the first order equation is as follows:
\begin{gather}
    (i\Omega+\frac12\Gamma)B_1+B_1(-i\Omega+\frac12\Gamma)+\Gamma=0,
\end{gather}
where $B_1=-I$ is the solution, where $I$ is identity matrix. Then $B
\approx -n_T I$ and it resembles the single-mode solution in App.~\ref{app1}, indeed substitution recovers that $B=-\frac{n_T}{n_T+1}I$ is full solution to the equation $R=0$.

\section{Eigenvalue equation}\label{appeigen}
We would like to investigate spectrum of diagonalized Liouvillian in the following form:
\begin{gather}
    \hat{L}^{(d)}\hat{\rho}=\lambda \hat{\rho}.
\end{gather}
One may introduce following notation as a shorthand:
\begin{gather}
    \hat{\rho}=\sum_{a,b,a_0,b_0}C_{a,b}^{a_0,b_0}|a\rangle_{1}\langle b|\otimes |a_0-a\rangle_{2}\langle b_0-b|\equiv \nonumber\\
    \equiv    \sum_{a,b,a_0,b_0}C_{a,b}^{a_0,b_0} (a,b,a_0,b_0),
\end{gather}
then
\begin{gather}
     \hat{L}^{(d)}\sum_{a,b,a_0,b_0}C_{a,b}^{a_0,b_0} (a,b,a_0,b_0)=\nonumber\\
     =\lambda^{a_0,b_0} \sum_{a,b,a_0,b_0}C_{a,b}^{a_0,b_0} (a,b,a_0,b_0)
\end{gather}
and
\begin{gather}
    \text{Tr} \Big(\hat{L}^{(d)}\sum_{a,b,a_0,b_0}C_{a,b}^{a_0,b_0} (a,b,a_0,b_0) (m,n,v,u) \Big)=\nonumber\\
   \text{Tr} \Big(\lambda^{a_0,b_0} \sum_{a,b,a_0,b_0}C_{a,b}^{a_0,b_0} (a,b,a_0,b_0) (m,n,v,u) \Big).
\end{gather}
Hence RHS of the latter equation simplifies as follows:
\begin{gather}
    \text{Tr} \Big(\lambda^{a_0,b_0} \sum_{a,b,a_0,b_0}C_{a,b}^{a_0,b_0} (a,b,a_0,b_0) (m,n,v,u) \Big)=\nonumber\\
    =\lambda^{u,v}C_{n,m}^{u,v},
\end{gather}
and LHS is as follows:
\begin{gather}
  \sum_{a,b,a_0,b_0} C_{a,b}^{a_0,b_0} \text{Tr}\Big(\big(\hat{L}^{(d)}(a,b,a_0,b_0)\big) (m,n,v,u) \Big)=\nonumber\\
 = \sum_{a,b} C_{a,b}^{u,v} (\hat{L}^{(d)})_{(a, n),(b, m)}^{u,v}.
  \end{gather}

\section{Linear approximation derivation}\label{appder}
Let us begin with
\begin{gather}
    e^{\hat{L}t}\approx e^{\hat{L}^{(0)}t}+n_T[\hat{\mathcal{K}}^{(+)}_{I},e^{\hat{L}^{(0)}t}]-n_T[\hat{\mathcal{K}}^{(-)}_{I},e^{\hat{L}^{(0)}t}].
\end{gather}

Then one may derive
\begin{gather}
    [\hat{\mathcal{K}}^{(+)}_{I}-\hat{\mathcal{K}}^{(-)}_{I},e^{\hat{L}^{(0)}t}]=\nonumber\\
    =\big(\hat{\mathcal{K}}^{(+)}_{I}-\hat{\mathcal{K}}^{(-)}_{I}-e^{\hat{L}^{(0)}t}(\hat{\mathcal{K}}^{(+)}_{I}-\hat{\mathcal{K}}^{(-)}_{I})e^{-\hat{L}^{(0)}t}\big)e^{\hat{L}^{(0)}t},
\end{gather}
then
\begin{gather}
    e^{\hat{L}^{(0)}t}(\hat{\mathcal{K}}^{(+)}_{I}-\hat{\mathcal{K}}^{(-)}_{I})e^{-\hat{L}^{(0)}t} =\nonumber\\
    =e^{-\hat{\mathcal{K}}^{(-)}_{I}} e^{\hat{L}^{(d)}t} e^{\hat{\mathcal{K}}^{(-)}_{I}} (\hat{\mathcal{K}}^{(+)}_{I}-\hat{\mathcal{K}}^{(-)}_{I}) e^{-\hat{\mathcal{K}}^{(-)}_{I}} e^{-\hat{L}^{(d)}t} e^{\hat{\mathcal{K}}^{(-)}_{I}}=\nonumber\\
    e^{-\hat{\mathcal{K}}^{(-)}_{I}} e^{\hat{L}^{(d)}t} (\hat{\mathcal{K}}^{(+)}_{I}+2\hat{\mathcal{K}}^{(0)}_{I}) e^{-\hat{L}^{(d)}t} e^{\hat{\mathcal{K}}^{(-)}_{I}},
\end{gather}
where
\begin{gather}
    e^{\hat{L}^{(d)}t} \hat{\mathcal{K}}^{(0)}_{I} e^{-\hat{L}^{(d)}t}=\hat{\mathcal{K}}^{(0)}_{I}\\
    e^{\hat{L}^{(d)}t} \hat{\mathcal{K}}^{(+)}_{I} e^{-\hat{L}^{(d)}t}=\hat{\mathcal{K}}^{(+)}_{U(t)},
\end{gather}
where 
\begin{gather}
    U(t)=e^{\big([-i\Omega,\ \cdot\ ]+\{-\frac{\Gamma}{2},\ \cdot\ \}\big)t}\cdot I=\nonumber\\
    = e^{\big(\overleftarrow{(-i\Omega-\frac{\Gamma}{2})}+\overrightarrow{(i\Omega-\frac{\Gamma}{2})}\big)t}\cdot I=\nonumber\\
    =e^{(-i\Omega-\frac{\Gamma}{2})t}\cdot e^{(i\Omega-\frac{\Gamma}{2})t}.
\end{gather}
In two-mode case explicit expression for $U(t)$ can be easily found by composition law of the group $SU(2)$ as corresponding exponentiated matrices ($
\pm i\Omega-\frac{\Gamma}{2}$) may be described in terms of Pauli matrices. Also
\begin{gather}
    e^{-\hat{\mathcal{K}}^{(-)}_{I}} \hat{\mathcal{K}}^{(+)}_{U(t)} e^{\hat{\mathcal{K}}^{(-)}_{I}}=\hat{\mathcal{K}}^{(+)}_{U(t)}-2\hat{\mathcal{K}}^{(0)}_{U(t)}+\hat{\mathcal{K}}^{(-)}_{U(t)},\\
    e^{-\hat{\mathcal{K}}^{(-)}_{I}} \hat{\mathcal{K}}^{(0)}_{I} e^{\hat{\mathcal{K}}^{(-)}_{I}}=\hat{\mathcal{K}}^{(0)}_{I}-\hat{\mathcal{K}}^{(-)}_{I}.
\end{gather}
Then
\begin{gather}
   [\hat{\mathcal{K}}^{(+)}_{I}-\hat{\mathcal{K}}^{(-)}_{I},e^{\hat{L}^{(0)}t}]=(\hat{\mathcal{K}}^{(+)}+2\hat{\mathcal{K}}^{(0)}+\hat{\mathcal{K}}^{(-)})_{I-U(t)} e^{\hat{L}^{(0)}t},
\end{gather}
and finally
\begin{gather}
    e^{\hat{L}t}\approx\Big(I+n_T(\hat{\mathcal{K}}^{(+)}+2\hat{\mathcal{K}}^{(0)}+\hat{\mathcal{K}}^{(-)})_{I-U(t)}\Big)e^{\hat{L}^{(0)}t}.
\end{gather}

\bibliography{bibliography1}

\end{document}